\documentclass[sigconf]{acmart}
\usepackage{rotating}
\usepackage{hyperref}
\usepackage{xcolor}
\AtBeginDocument{%
  \providecommand\BibTeX{{%
    \normalfont B\kern-0.5em{\scshape i\kern-0.25em b}\kern-0.8em\TeX}}}

\newcommand{\Y}[1]{{#1}}

\copyrightyear{2024}
\acmYear{2024}
\setcopyright{acmlicensed}\acmConference[CHI '24]{Proceedings of the CHI Conference on Human Factors in Computing Systems}{May 11--16, 2024}{Honolulu, HI, USA}
\acmBooktitle{Proceedings of the CHI Conference on Human Factors in Computing Systems (CHI '24), May 11--16, 2024, Honolulu, HI, USA}
\acmDOI{10.1145/3613904.3642134}
\acmISBN{979-8-4007-0330-0/24/05}

\usepackage{float}
\begin{document}

\title{Shaping Human-AI Collaboration: Varied Scaffolding Levels in Co-writing with Language Models}


\author{Paramveer S. Dhillon}
\authornote{Corresponding Author}
\affiliation{%
  \institution{University of Michigan}
  \city{Ann Arbor}
  \state{MI}
  \country{USA}}
  \email{dhillonp@umich.edu}

\author{Somayeh Molaei}
\affiliation{%
  \institution{University of Michigan}
  \city{Ann Arbor}
  \state{MI}
  \country{USA}}
  \email{somayeh.molaei@gmail.com}

\author{Jiaqi Li}
\affiliation{%
  \institution{University of Michigan}
  \city{Ann Arbor}
  \state{MI}
  \country{USA}}
  \email{ellali@umich.edu}

\author{Maximilian Golub}
\affiliation{%
  \institution{University of Michigan}
  \city{Ann Arbor}
  \state{MI}
  \country{USA}}
  \email{maxgolub@umich.edu}

\author{Shaochun Zheng}
\affiliation{%
  \institution{UC San Diego}
  \city{La Jolla}
  \state{CA}
  \country{USA}}
  \email{zhengsc@ucsd.edu}

\author{Lionel P. Robert}
\affiliation{%
  \institution{University of Michigan}
  \city{Ann Arbor}
  \state{MI}
  \country{USA}}
  \email{lprobert@umich.edu}
  
\renewcommand{\shortauthors}{Dhillon et al.}

\begin{abstract}
 Advances in language modeling have paved the way for novel human-AI co-writing experiences. This paper explores how varying levels of scaffolding from large language models (LLMs) shape the co-writing process. Employing a within-subjects field experiment with a Latin square design, we asked participants (N=131) to respond to argumentative writing prompts under three randomly sequenced conditions: no AI assistance (control), next-sentence suggestions (low scaffolding), and next-paragraph suggestions (high scaffolding). {\bf Our findings reveal a U-shaped impact of scaffolding on writing quality and productivity (words/time).} While low scaffolding did not significantly improve writing quality or productivity, high scaffolding led to significant improvements, especially benefiting non-regular writers and less tech-savvy users. No significant cognitive burden was observed while using the scaffolded writing tools, but a moderate decrease in text ownership and satisfaction was noted. Our results have broad implications for the design of AI-powered writing tools, including the need for personalized scaffolding mechanisms.
\end{abstract}





\maketitle

\section{INTRODUCTION}
Recent advances in large language models (LLMs) like GPT-3~\cite{brown2020}, Gopher~\cite{rae2021scaling}, and PaLM~\cite{chowdhery2022palm} suggest exciting new possibilities for AI-powered writing assistance. LLMs can now generate fluent, coherent text spanning multiple sentences or even paragraphs with minimal input. This enables richer interactions than traditional word-level auto-completion~\cite{Bhat2023}. Rather than just accelerating typing, LLMs can actively collaborate as co-writers that provide meaningful suggestions tailored to the context of the piece~\cite{Lee_2022}. The prospect of an AI assistant that fluidly contributes ideas, expands on concepts, or even builds on the human writer's tone and style is highly compelling. Such technology could enhance productivity for a wide range of users, including students and professionals. It also creates new avenues for aiding struggling writers or personalizing support based on individual needs and backgrounds~\cite{Nazari2021}. If thoughtfully implemented, AI co-writing has immense potential to augment human creativity and expression~\cite{Jakesch2023,Mirowski2023}. However, the technology remains in its infancy, and extensive research on human-centered interactions is crucial as generative AI capabilities rapidly advance. For instance, it is unclear how to best structure the human-AI collaboration for co-writing~\cite{amershi2014power}. Understanding the right balance of AI assistance is critical for realizing the benefits of this emerging technology while maintaining human agency and control in the writing process~\cite{amershi2019guidelines}. We need further research to provide evidence-based guidance on integrating AI co-writers in a manner conducive to human satisfaction and productivity.

Minimal AI assistance, such as simple word suggestions or basic grammatical corrections, may not offer substantial benefits to writers. This is particularly true for experienced writers who may already possess a rich vocabulary and a strong grasp of syntax. For instance, a writer crafting a complex technical report may find minimal AI assistance not much different from basic spell-checkers, offering little in the way of enhancing the depth or coherence of the content~\cite{buschek2021impact}. On the other end of the spectrum, excessive AI scaffolding, such as full-paragraph generation, carries its own set of challenges. While it may expedite the writing process, it can also deter the writer from critically engaging with the subject matter, reducing the depth of their understanding and their emotional investment in the project~\cite{Yang2022,cheng2022mapping}. For example, a student working on an essay might lean heavily on AI-generated paragraphs but could end up not fully grasping the nuanced arguments they are supposed to be making. It is critical for human writers to have adequate AI help and to retain ownership of their own text, thus justifying further research into human-AI scaffolding.

In the context of AI-supported writing, the scaffolding dynamics are highly complex. Expert writers may eschew AI assistance, fearing it could stifle their creativity or impose a ``cookie-cutter'' structure on their unique style~\cite{Savvas2023}. Conversely, non-experts or those experiencing writer's block might find initial AI suggestions invaluable to kick-start their writing process, editing and refining AI-generated content to better align it with their own voice and objectives~\cite{Gilburt2023}. Similarly, novice writers might benefit more from comprehensive AI support, where low-level suggestions, like individual sentence prompts, may be inadequate ~\cite{Bhat2023}. For them, low-level suggestions such as individual sentence prompts might prove ineffective, akin to giving a beginner carpenter only a hammer when they need an entire toolbox. High-level scaffolding can provide a more structured framework, akin to giving carpenters a blueprint to follow, thereby reducing the cognitive load and making the writing task more manageable.
\Y{ All these use-cases underscore the importance of finely tuning the degree of AI intervention to match the writer's expertise and needs, highlighting a crucial aspect of human-AI interaction in writing. This necessity for careful calibration also emphasizes the need for AI systems to be adaptable to various user requirements and writing stages~\cite{storch2005collaborative}. Consequently, a deep understanding of the intricate dynamics of human-AI co-writing becomes essential. A detailed analysis of contextualized, incremental scaffolds that mirror adaptive human tutoring is vital to grasp these dynamics~\cite{chen2019sentiment}. Such an understanding is fundamental in designing AI-assisted writing tools tailored to the nuanced needs of different user groups.}

\Y{Our work seeks to address these research gaps by centering itself on the evolving role of AI in augmenting the human writing process. In particular, our work asks the key research question: {\bf What granularity of AI support is most effective for writing assistance, and for which group of writers is it most beneficial?} We investigate this question by considering {\it two different levels of AI assistance, namely sentence and paragraph suggestions.} This choice is grounded in the fundamental structure of language writing where sentences form the basic building blocks of coherent thought, and paragraphs represent larger units of ideas or arguments~\cite{lunsford2016everything}. By exploring these natural levels of composition, we align AI assistance with the inherent process of human writing, allowing for more intuitive and effective collaboration. Hence, our work responds to a gap in current research: {\it while the use of AI in writing is well-documented, there's limited insight into how varying intensities of AI input influence user experience and writing outcomes.} The insights gained from our study can be pivotal for the design of AI writing tools that are truly user-centric. By understanding how different levels of AI input can aid or hinder the writing process, we can tailor AI tools to better match user needs, enhancing their overall writing experience. This aligns with the forefront of AI-HCI research, which seeks to harmonize AI capabilities with human skills and preferences. Our findings provide practical guidelines for developing AI tools that not only improve writing quality and productivity but also maintain user satisfaction and a sense of ownership over their work.}

To rigorously examine these considerations, we employ a within-subjects field experiment designed with a Latin square arrangement (to mitigate order effects)~\cite{richardson2018use}. This design ensures that each participant is exposed to every treatment condition, thereby enhancing the internal validity of the study. In this experiment, participants are tasked with crafting responses to a series of carefully selected argumentative writing prompts. {\bf We evaluate the effects of AI-based scaffolding under three carefully controlled conditions: a baseline condition with no AI assistance, a low-level scaffolding condition offering next-sentence suggestions, and a high-level scaffolding condition that provides more comprehensive next-paragraph suggestions.}

This multi-tiered approach not only permits us to compare the relative effectiveness of varying levels of AI-based scaffolding but also provides a granular view of how different groups\textemdash be they novices, intermediates, experts, tech-savvy, or non-tech users\textemdash respond to each level of support. By isolating these variables, our work aims to reveal fine-granular insights into the optimal balance between AI assistance and human agency in writing tasks as they pertain to different types of people. This contributes to the theoretical understanding and practical applications of AI-assisted writing.

\Y{{\bf Our empirical findings reveal a complex, U-shaped relationship between the level of AI scaffolding in writing and the metrics of writing quality and productivity (words/time). Notably, low-level scaffolding\textemdash defined as next-sentence suggestions\textemdash did not significantly enhance these outcomes.} This observation aligns with the scaffolding theory posited by Wood et al.~\cite{wood1976role}, which emphasizes the importance of providing support that is appropriate to the learner's (or in this case, writer's) current level of proficiency. Sentence-level guidance, while intended to assist, may instead be too rudimentary and disruptive, compelling the writer to frequently oscillate between their own writing process and the AI's suggestions. This disruption can impede the smooth flow of ideas and concentration, resonating with Vygotsky’s~\cite{vygotsky1978development} concept of the Zone of Proximal Development, where the support provided should ideally challenge but not overwhelm the learner.}

\Y{{\bf Conversely, high-level scaffolding\textemdash exemplified by next-paragraph suggestions\textemdash resulted in marked improvements in writing quality and productivity.} This finding is in line with the principles of effective scaffolding, where support is provided in a manner that extends the writer’s current capacity~\cite{wood1976role}. Paragraph-level suggestions offer a more holistic framework that writers can adapt or modify, effectively reducing cognitive load and facilitating a more focused writing experience. These benefits were particularly pronounced among non-regular writers and those less familiar with advanced technology, supporting the notion that adaptive writing tools are especially advantageous for these groups.}

\Y{{\bf However, this enhanced writing experience was accompanied by a notable caveat: a moderate decrease in perceived text ownership and overall satisfaction among users of scaffolded writing tools.} This finding echoes the concerns raised in scaffolding literature about the balance between providing support and fostering independence~\cite{pea2018social}. It highlights the complexities inherent in the design of AI-powered writing assistants, emphasizing the need for creating adaptive, personalized scaffolding mechanisms that are sensitive to the user's evolving needs and proficiency levels. Such an approach would not only enhance writing productivity and quality but also maintain a sense of ownership and satisfaction in the writing process, aligning with the core objectives of scaffolding as a facilitative tool in learning and development.}

Given the rapid advances in language model capabilities, our findings offer timely insights into crafting writing tools that enhance the user experience, aligning with key Human-Computer Interaction (HCI) considerations around human-AI collaboration~\cite{amershi2019guidelines}. While language models grow more sophisticated, keeping human needs at the center remains imperative. Our research constitutes an early but significant step toward this human-centric approach of leveraging AI to facilitate richer expressions of creativity, knowledge, and ideas.

\section{RELATED WORK}

\subsection{AI-based writing tools \& Human-AI collaboration}

Tools facilitating writing collaboration between humans and AI have been extensively studied in the literature. Initially, these tools employed straightforward algorithms aimed at enhancing quality of life through minor improvements, such as spelling error detection using deterministic approaches~\cite{Peterson2023}. Gradually, the technology advanced to support more complex functions, including automated text correction, phrase-level completion~\cite{Xiao2014, Mingrui2019, Arnold2016}, the provision of interactive thesauri~\cite{Gero2019}, metaphor generation~\cite{Mermaid2021}, and functionalities like Gmail's Smart Reply and Smart Compose~\cite{Chen2019, Kannan2016}, along with text summarization applications~\cite{cheng2022mapping}. Despite these advancements, the early iterations of such tools were mainly limited to generating predictive text or adjusting text already written by users. This limitation has been linked to potential decreases in user satisfaction and writing efficiency, as highlighted by various studies~\cite{Girish2016, Banovic2019, Quinn2016}.

Due to their restrictive nature, these technologies were succeeded by generative deep learning models that can produce longer responses that are more open-ended and inventive than their earlier counterparts~\cite{Vaswani2017, winata2021}. 
This shift redefines AI as an active collaborator, expanding its use in diverse contexts like story writing~\cite{Singh2022, Bhat2023, Chung2022, Stefnisson2018, Kreminski2020,yuan2022wordcraft} and creative tasks~\cite{Clark2018, Gero2019Two, Yang2022, Lee_2022, Gabriel2015}. These adaptable models support specialized tasks such as journalistic ideation~\cite{Savvas2023}, visual storyboarding~\cite{Chung2022}, screenplay co-creation~\cite{Mirowski2023}, emotional composition~\cite{Peng2020}, crafting help requests~\cite{Hui2018}, and generating scientific paper ideas~\cite{gero2021sparks}.

\Y{Several previous studies have laid the groundwork for understanding the interaction between humans and AI in writing contexts. For instance, Kreminski et al.~\cite{Kreminski2020}, Stefánsson’s Mimisbrunnr~\cite{Stefnisson2018}, and the WordCraft paper~\cite{yuan2022wordcraft} have all explored how generative AI can assist in organizing thoughts or actively contribute to generating text. Mimisbrunnr, in particular, demonstrates AI's capability in setting up a narrative framework, enabling authors to co-write stories with AI by sharing ideas and developing overarching themes.}

\Y{Our research builds upon this literature in several ways. First, we introduce a novel approach to evaluating co-authored texts. We have developed a set of extensive quantitative metrics, which go beyond the scope of earlier studies. Our approach incorporates both human evaluation and automated text evaluation tools like TAACO\footnote{\url{https://www.linguisticanalysistools.org/taaco.html}} and TAALES\footnote{\url{https://www.linguisticanalysistools.org/taales.html}}. These tools offer objective assessments of text cohesion and linguistic sophistication, respectively, providing a more comprehensive understanding of the quality of co-authored texts.} \Y{Second, our work differentiates itself from recent studies, such as the Co-Author paper by Lee et al.~\cite{Lee_2022}, by providing a detailed hybrid evaluation of the written text. Finally, our evaluation-related contributions are complemented by our work's examination of multiple levels of AI scaffolding. We analyze how varying degrees of AI assistance influence the writing process, offering insights into the optimal integration of AI in enhancing human creativity. This aspect of our study provides a deeper exploration into the dynamics of AI-human collaboration, focusing not just on the end product but also on the process itself.}

In addition to the aforementioned work on developing AI-assisted writing technologies, an increasing number of studies within the field of Human-Computer Interaction (HCI) are focusing on the broader implications and utility of AI-based co-authoring tools. This includes studies on how these tools change the dynamics of writing, revealing the influence of technology on the creative process~\cite{Bhat2023}. There is also a growing interest in how users of different language proficiencies utilize these tools, highlighting variations in their effectiveness~\cite{Buschek2021}. Research is also being conducted to understand how biased language models might shape user writing and viewpoints~\cite{Jakesch2023}. Finally, Fugener et al.~\cite{Fugener2022} have recently sparked interest in a line of work that studies cognitive challenges involved in productive delegation in Human-AI collaboration. \Y{Our work advances this direction of research by offering fresh perspectives on the nuanced role of AI in writing, providing insights into how varying AI support levels influence emotional and cognitive investment, task efficiency, and user control in argumentative writing. It highlights the differential impacts of AI assistance on users with varied writing expertise and technology familiarity, emphasizing the importance of striking a balance in AI assistance for enhancing writing quality while maintaining user engagement and authorship.}

\subsection{Scaffolding Strategies in Education Research}

\Y{Scaffolding, a well-studied concept in educational psychology, refers to the temporary instructional support given to students to enhance their task performance. The foundational theory of scaffolding, introduced by Wood et al.~\cite{wood1976role}, posits that such support enables learners to acquire skills and knowledge that would otherwise be difficult to attain independently. Complementing this, Storch's work~\cite{storch2005collaborative} offers critical insights into collaborative writing and the role of tutoring in problem-solving, emphasizing the adaptability of scaffolding to each student's skill level and individual needs.}

\Y{ Building on these fundamental theories, recent studies have explored the effects of various scaffolding strategies in different contexts, particularly focusing on English as a Foreign Language (EFL) writing proficiency. These studies have employed a range of strategies, including high- versus low-structure scaffolds, portfolio-based assessment, teacher- and peer feedback, process writing, self-revision, and mediated learning~\cite{baleghizadehSocioculturalPerspectiveSecond2011, obeiahEffectPortfolioBasedAssessment2016, gholamipasandPeerScaffoldingEFL2017, hanjaniCollectivePeerScaffolding2019, sumarnoEffectsEdmodoAssistedProcess2019, sulindraScaffoldingUsedWriting2018, khojastehImpactMediatedLearning2021}. Collectively, this body of work demonstrates that well-aligned scaffolding can significantly enhance writing proficiency.}

\Y{ Our work contributes to this literature by introducing an innovative form of scaffolding\textemdash AI-based guidance, with a particular focus on next-sentence and next-paragraph suggestions. Essentially, we vary the granularity of AI-assistance, reflecting the incremental and contextual support that characterizes effective human tutoring in writing. Our work not only adds a new dimension to the existing scaffolding literature but also deepens our understanding of the potential and limitations of AI in assisting the writing process~\cite{gero2019metaphoria,peng2020exploring}.}

\section{METHODS}

\begin{figure*}[h]
  \centering
  \includegraphics[width=\textwidth]{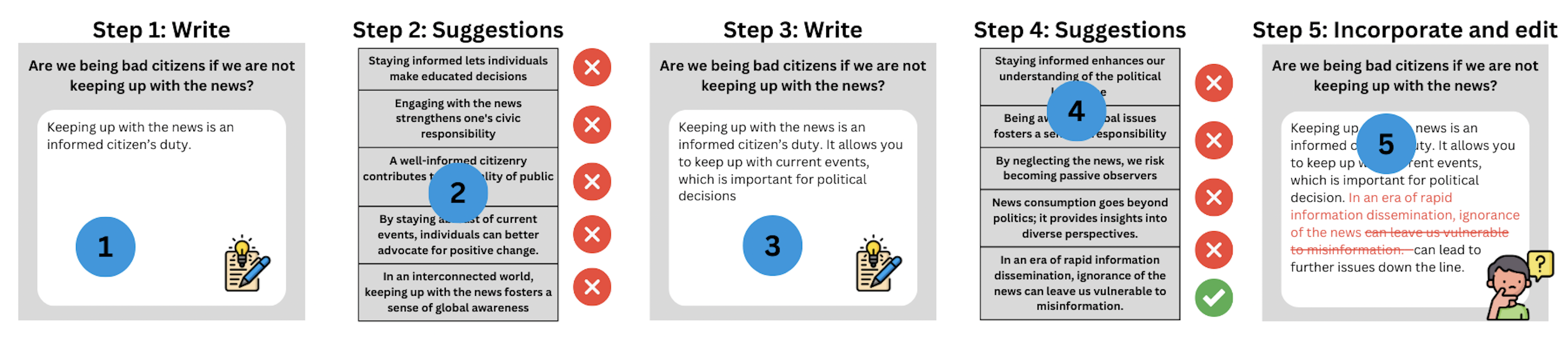} 
  \caption{\Y{User Interaction with our interface: 1) The user writes down a sentence on their own. 2) The user presses the tab key to generate five suggestions but rejects all of them. 3) The user writes more on their own. 4) The user presses the tab key to generate suggestions, but this time selects the last one. 5) The generated text is added to the end of the user’s text, and the user edits it as they see fit.}
  \Description{Figure 1 shows an example of a potential process by which a participant can interact with our interface: Step one: The user writes down a sentence on their own in our writing box based on the prompt. Step two: The user presses the tab key to generate five suggestions but rejects all of them. Step three: The user continues writing on their own. Step four: The user presses the tab key again to generate suggestions, and this time selects the last suggestion. Step five: The generated text is added to the end of the user's text, and the user edits it as they see fit.} 
  \label{fig:overview_process_fig}
 }
\end{figure*}

We conduct a field experiment to explore how individuals engage with an AI-based collaborative writing tool and how varying levels of scaffolding from large language models (LLMs) shape the co-writing process. A screened participant pool (N=131) was invited to complete argumentative writing tasks with varying levels of AI assistance.\footnote{\Y{Many selected participants were unable to complete our main study due to scheduling conflicts.}} Our custom tool systematically collected data, capturing user inputs and diverse evaluation metrics (cf. Figures~\ref{fig:overview_process_fig}, \ref{fig:overview_interface_fig}).

\subsection{Experiment design}

\subsubsection{Pre-screening task:}
We designed a pre-screening writing task based on a TOEFL argumentative prompt to evaluate participants, detailed in Appendix~\ref{appendix:Pre_screening_HAI_Collaborative_Writing_Task}. This task assesses participants' ability to summarize and contrast perspectives within a 150-225 word limit, emphasizing writing quality and argument effectiveness. The pre-screening task was completed online by the participants with a 40-minute time limit. This approach standardized the conditions under which participants completed their tasks, facilitating a fair comparison of their abilities. Participants' responses were scored on a 1-5 scale by two evaluators using the TOEFL rubric (Appendix~\ref{appendix:Pre_screening_Task_Rubric})~\cite{ets2023}. Finally, we invited the participants who scored 2 or higher to our main experiment. We chose a relatively low threshold score of 2 as we aimed to investigate how the degree of AI assistance in writing would impact individuals across a spectrum of writing skills, including both proficient and less proficient writers.

\subsubsection{Main Experiment:}
\begin{figure*}[h]
  \centering
  \includegraphics[width=0.94\textwidth]{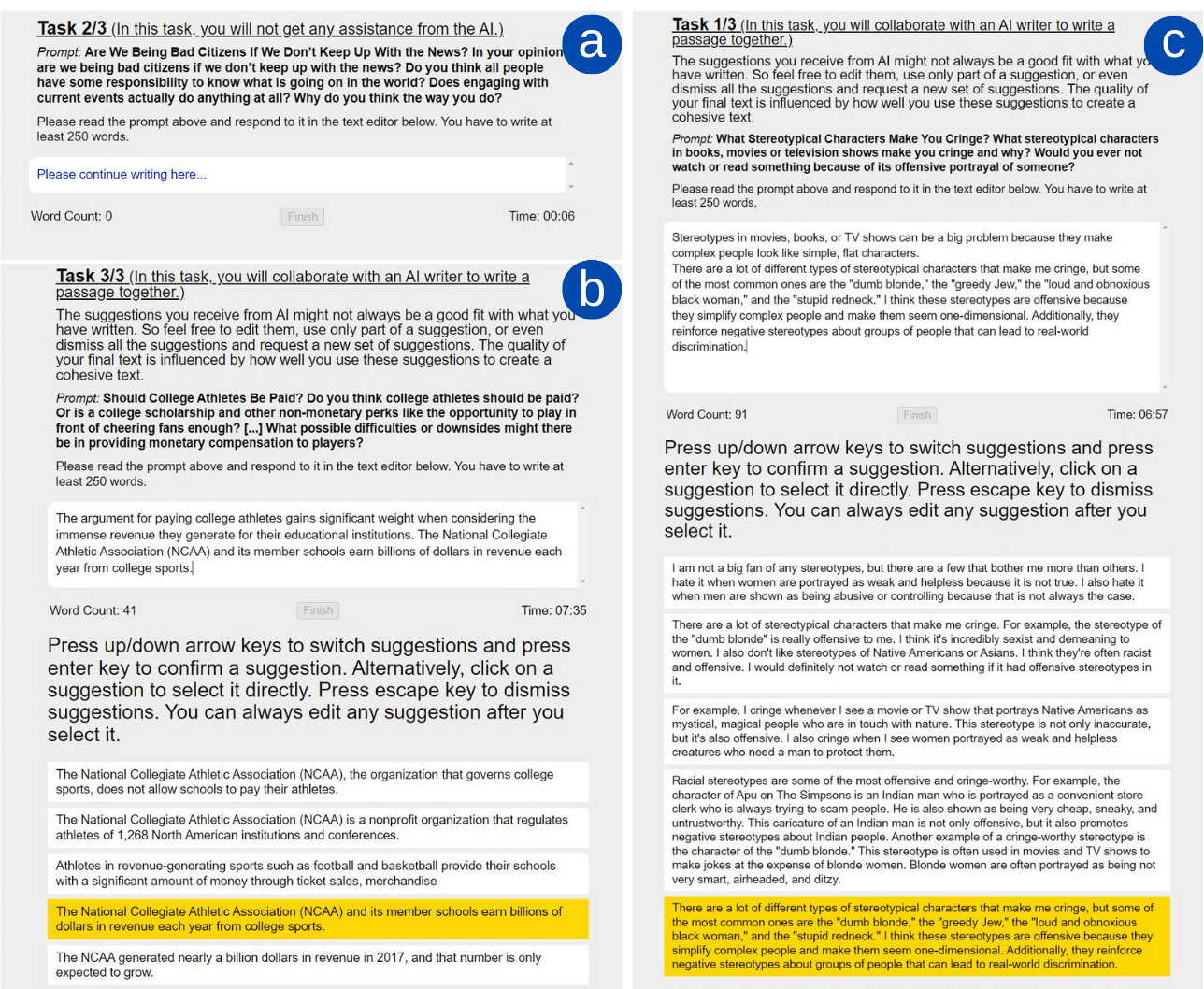} 
  \caption{\Y{Our custom AI co-writing interface and the three treatment conditions: a) No AI assistance (control), b) Sentence Mode (low scaffolding) and c) Paragraph Mode (high scaffolding).}}
 \Description{Figure 2 shows three screenshots of our custom-built interface under three treatment conditions: No AI assistance (control), Sentence Mode (low scaffolding), and Paragraph Mode (high scaffolding). Each interface consists of a task number, an AI-assistance mode description, a writing prompt, and a word count requirement. A real-time word count is displayed at the bottom left of the writing text box, and the time spent on the task is shown at the bottom right. A "Complete" button becomes active when the written answer reaches at least 250 words. In Sentence and Paragraph Modes, up to five AI-generated suggestions appear below the text box after pressing the Tab key. The currently selected suggestion is highlighted. Instructions on how to use AI suggestions effectively are also provided.} 
  \label{fig:overview_interface_fig}
\end{figure*}

\paragraph{Procedure for Live Virtual Writing Sessions:}
To ensure the integrity of our main experiment, we held multiple live virtual writing sessions through Zoom, each accommodating between 1 and 10 participants. During these sessions, participants were provided a link directing them to the custom-built writing tool where they would complete their writing tasks (please see Figure~\ref{fig:overview_interface_fig}). Each session was overseen by a facilitator who introduced the instructions, responded to participants' queries, and provided assistance throughout the process. To minimize interference among participants, each individual was assigned to a separate breakout room to complete the task. Additionally, participants were required to activate their cameras and share their screens, a measure implemented to confirm that they were completing the tasks independently. This procedural setup was designed to create a controlled yet flexible environment for capturing reliable participant writing data.

\paragraph{Levels of Scaffolding (Treatment Conditions):} We designed the writing tasks with three different levels of scaffolding:
\begin{itemize}
    \item \underline{no AI assistance (control):} In this condition, participants received no assistance from the AI, meaning they could not access any AI-generated suggestions.
    \item \underline{next-sentence suggestions (low scaffolding):} Participants could obtain sentence-level suggestions from the AI in this mode.
    \item \underline{next-paragraph suggestions (high scaffolding):} Participants could obtain paragraph-level suggestions from the AI in this mode.
\end{itemize}

\paragraph{Experiment Design:} 
Our study utilized a within-subjects design to efficiently examine AI scaffolding effects, employing a Latin Square design~\cite{richardson2018use} to evenly distribute scaffolding conditions and minimize order biases. This design allowed each participant to experience all three conditions, enhancing the understanding of each condition's impact and revealing how the sequence of AI assistance affects participant behavior and perceptions. 
Interestingly, this approach also inadvertently facilitated an investigation into the positional effects of AI assistance on participant responses. While not the primary focus of our initial research question, this aspect of the study offered valuable insights into how the sequence of exposure to AI scaffolding conditions could influence participant behavior and perceptions, adding an unexpected dimension to our findings. By tracking participants across different AI assistance conditions, we gain valuable insights into how the placement and level of AI support affects their writing process.

Participants were randomly assigned to one of three ordered sequences (37 in sequence 1, 53 in sequence 2, 41 in sequence 3), each experiencing all conditions with unique prompts to prevent response learning, thereby enhancing study validity.
\begin{itemize}
    \item \underline{Sequence 1:} no AI assistance (control) $\rightarrow$ next-sentence suggestions (low scaffolding), $\rightarrow$ next-paragraph suggestions (high scaffolding).
    \item \underline{Sequence 2:} next-paragraph suggestions (high scaffolding) $\rightarrow$ no AI assistance (control) $\rightarrow$ next-sentence suggestions (low scaffolding).
    \item \underline{Sequence 3:} next-sentence suggestions (low scaffolding) $\rightarrow$ next-paragraph suggestions (high scaffolding) $\rightarrow$ no AI assistance (control).
\end{itemize}
\paragraph{Writing Prompts:} We utilized 10 argumentative writing prompts originally selected by the CoAuthor study~\cite{Lee_2022} from a set provided by The New York Times~\cite{network300QuestionsImages2021}. These prompts were chosen to offer a range of accessible and balanced topics, aligning well with our research goals. A complete list of the prompts can be found in Appendix~\ref{appendix:main_study_argumentative_prompts}.

\paragraph{Procedure for AI-Assisted Writing Tasks}
After giving informed consent, participants completed a pre-task survey (see Appendixx~\ref{appendix:survey_interface}) that gathered demographic information. Then, they transitioned to the main writing task. The main writing task was split into three sub-tasks corresponding to different scaffolding levels: no AI assistance (control), next-sentence suggestions (low), and next-paragraph suggestions (high), detailed in Figure~\ref{fig:overview_interface_fig}. Each sub-task required responding to a writing prompt in a text editor, aiming for a minimum of 250 words. For AI-assisted conditions, suggestions were available on request by pressing the ``Tab'' key. Upon completing each sub-task, participants were prompted to complete two post-task surveys  described in Section 3.4 (and Appendix~\ref{appendix:survey_interface}). The complete set of post-task survey questions are provided in Table~\ref{table:postsurvey-questions}.

\subsection{The Custom-Built Collaborative Writing AI Tool}

We created a custom collaborative writing AI Tool for the experiment. We utilized Facebook's React and the Lexical text editor framework~\cite{Lexical2023} owing to its reliability. We enriched this baseline framework with necessary user interactions, features, and surveys for our study, and integrated the GPT-3 DaVinci engine for text completion.

\subsubsection{User Interface}
Our interface consisted of a text editor that was pre-configured with a writing prompt and equipped with keyboard shortcut descriptions for various features, as shown in Figure~\ref{fig:overview_interface_fig}. Standard text interactions like typing, selection, editing, and deletion were fully supported, along with cursor navigation via keyboard or mouse (details in Appendix B). A timer was displayed at the bottom-right corner of the screen and a word count meter was situated at the bottom-left corner. Once the final text reached the 250-word benchmark, the ``Finish'' button would activate, allowing participants to proceed to the next task.

\subsubsection{User Interactions}
In the two experimental conditions involving AI suggestions\textemdash specifically, Sentence Mode and Paragraph Mode\textemdash participants were permitted to request up to five suggestions from the AI by pressing the `Tab' key. This action triggered a request to the GPT-3's API, which included both the prompt and the participant's written content up to that point. Subsequently, a panel below the text editor displayed a list of five suggestions on the participant's screen. Participants could select, confirm, or dismiss a suggestion using either the mouse or the keyboard. By default, the first suggestion was pre-selected and appeared as a preview in the text editor. Participants could navigate through the list of suggestions using the up and down arrow keys or by hovering over a button in the selection panel to make a direct selection. To confirm a chosen suggestion, participants could either press the `Enter' key or click the corresponding suggestion button. Once confirmed, the suggestion was automatically inserted into the text, and the selection panel was hidden. A suggestion was considered `accepted' when selected by the participant from the displayed list and was inserted automatically at the cursor's current location. To dismiss all suggestions, participants could press the `Escape' key, resulting in the removal of the previewed text and the hiding of the selection panel. Subsequently, they could request a new set of suggestions. Participants also had the flexibility to modify or partially adopt any selected suggestions. Upon completing a writing session, participants were required to click the `Finish' button to proceed. Advancement to the subsequent steps was restricted until participants had generated a minimum of 250 words. A full overview of our interfaces is illustrated in Figure~\ref{fig:overview_interface_fig}.

\Y{We used the parameters described in Table~\ref{table:model_parameters} in Appendix when sending a query to the GPT-3 API. We set a 30-token limit for sentence-based suggestions to keep them concise and relevant, ensuring they fit seamlessly into the user's existing text. This length mirrors typical sentence structures, aiding in natural integration. Similarly, we set a maximum length of 150 tokens for the paragraph mode.\footnote{We experimented with a few different values for maximum number of tokens in paragraph mode (100, 150, 200, 250), and we chose 150 since it resulted in more consistent outputs.} Sometimes, GPT-3 returned an empty string, which was promptly removed from the output. Following the lead of previous work~\cite{Lee_2022}, each user was provided five suggestions to choose from. These decisions were made to enhance the tool's practicality and user-friendliness in the writing process, aiming to create a seamless and efficient interface for users to interact with AI-generated content.} 

\begin{table*}[htbp]
  \begin{small}
  \begin{tabular}{clp{6cm}l}
    \toprule
    ID & Survey Category & Question & Scale \\
    \midrule
    1 & Mental Demand & How much mental and perceptual activity was required (e.g., thinking, deciding, calculating, etc.)? & Low to High \\
    2 & Physical Demand & How much physical activity was required (e.g., typing, staring at the monitor, etc.)? & Low to High \\
    3 & Temporal Demand & How much time pressure did you feel due to the rate or pace at which the task elements occurred? & Low to High \\
    4 & Effort & How hard did you have to work (mentally and physically) to accomplish your level of performance? & Low to High \\
    5 & Performance & How successful do you think you were in accomplishing the goals of the task set by the experimenter (or yourself)? & Good to Poor \\
    6 & Frustration & How insecure, discouraged, irritated, stressed, and annoyed were you during the task? & Low to High \\
    \midrule
    7 & AI Collaboration & You collaborated with an AI in this task. How satisfied are you with the output of this collaboration, i.e., the final response to the prompt question? & Not Satisfied to Very Satisfied \\
    8 & AI Collaboration & You collaborated with an AI on this task. How much ownership do you feel over the output of this collaboration, i.e., the final response to the prompt question? & No Ownership to Full Ownership \\
    9 & AI Collaboration & Has your opinion on the topic of the prompt been influenced by the text responses provided by the AI? & No Influence to Heavy Influence \\
    10 & AI Collaboration & How difficult was the given prompt to write about? & Easy to Difficult \\
    11 & AI Collaboration & How likely is it that you will use AI for writing in the future? & Not Likely to Very Likely \\
    \bottomrule
  \end{tabular}
  \end{small}
   \caption{Post-Task Survey Questions}
  \label{table:postsurvey-questions}
\end{table*}

\subsection{Participant Recruitment}
\Y{We initiated the recruitment process on January 17, 2023, and continued through August 8, 2023. Participants were sourced from the health-system of a large public university in the USA. This diverse participant pool included not just healthcare workers but also administrative and support staff, as well as some former or current patients. This group, encompassing around 50,000 individuals, represented various socio-economic, ethnic, and professional backgrounds, significantly reducing the risk of selection bias.}

\Y{In selecting our participant pool, we intentionally avoided online platforms like Amazon Mechanical Turk or Prolific. Our primary concern was the potential compromise in data quality and representativeness often associated with these platforms, as highlighted in recent studies~\cite{veselovsky2023artificial,douglas2023data,hauser2019common}.\nocite{dhillon11a} Additionally, the nature of our study necessitated meticulous monitoring of participant engagement to ensure data integrity and prevent the use of AI tools, like ChatGPT, in their responses\textemdash a level of oversight that would have been challenging to achieve with remote participants on platforms like Mechanical Turk or Prolific. To this end, we conducted over 100 hours of live Zoom sessions, allowing us to closely observe participants during the writing tasks. This rigorous monitoring approach was instrumental in providing reliable, high-quality data and was essential to maintaining the internal validity of our study.}

\Y{We posted our study on the health-system platform, and interested participants could sign up if they met the inclusion criteria. Initially, 890 people expressed their interest in participating. However, only 453 responded to our invitation for a screening task; the others either did not reply or had to cancel. Upon expressing interest, we contacted these individuals to participate in a pre-screening task. Post-screening, we selected participants who put in some ``effort'' in the writing task, as evidenced by scoring atleast 2 out of 7 from two independent raters. This criterion reduced our pool to around 200 people. However, due to scheduling conflicts and unforeseen issues, our final participant count for the main study was 131.}

\Y{All participants were from the United States, at least 18 years of age, and came from potentially diverse backgrounds. Compensation was provided for both the pre-screening and the main writing tasks. Specifically, participants were given a fixed rate of \$15 for the pre-screening task. For the main writing task, participants received a base compensation of \$10, with additional bonuses ranging from \$1 to \$9 (in increments of 2) based on the quality of their writing, amounting to a maximum total compensation of \$19.\footnote{The research received approval from the Institutional Review Board affiliated with the first author's institution (HUM\# 00221115).} The demographic summary of the participants is presented in Table~\ref{table:demographic_breakdown}.}

\begin{table*}[h]
  \begin{small}
  \begin{tabular}{lllp{0.01cm} lllp{0.01cm}}
    \toprule
    
    & Category & \% & & & Category & \% & \\
    \midrule
    Age & 18-25 years & 21.7\% & & First Language & English & 93.0\% & \\
        & 25-35 years & 24.0\% & &               & Non-English & 7.0\% & \\
        & 35-45 years & 15.5\% & & & & & \\
        & 45-55 years & 10.1\% & & & & & \\
        & 55+ years & 28.7\% & & & & & \\
    \midrule
    Gender & Female & 75.2\% & & English Proficiency & Native Speaker & 77.5\% & \\
           & Male & 22.5\% & &                    & Fluent in English & 21.7\% & \\
           & Non-binary & 2.3\% & &              & Not Fluent in English & 0.8\% & \\
    \midrule
    Education & Graduate degree & 51.9\% & & Writing Expertise & Professional Writer & 5.4\% & \\
              & College degree & 38.8\% & &                  & Writes Regularly & 62.8\% & \\
              & High-school & 9.3\% & &                   & Does Not Write Regularly & 31.8\% & \\
    \midrule
    Ethnicity & Hispanic/Latino & 3.1\% & & Prior Use of Assistive  & No & 31.0\% & \\
              & Not Hispanic/Latino & 95.3\% & &    Technology in Writing   & Basic Usage & 58.9\% & \\
              & Prefer not to say & 1.6\% & &                & Advanced Writing Assistants & 10.1\% & \\
    \midrule
    Race & White & 81.4\% & & Attitudes Toward & Enjoys Working with AI & 17.8\% & \\
         & American Indian/Alaska Native & 0.8\% & & Working with AI Tools  & Would Like to Collaborate & 51.9\% & \\
         & Asian & 11.6\% & &                               & Finds it Fun to Work with AI & 30.3\% & \\
         & Black/African American & 2.3\% & & & & & \\
         & Prefer not to say & 3.9\% & & & & & \\
    \bottomrule
  \end{tabular}
  \end{small}
    \caption{\Y{Comprehensive Demographic and Profile Information of Participants (N=131): This table details participants' demographics, language proficiency, writing expertise, use of assistive technology, and attitude towards AI tools.}}
  \Description{A comprehensive table detailing participants' demographic information, language proficiency, writing expertise, assistive technology usage, and attitudes towards AI tools. The age distribution is as follows: 18-25 years:  21.7\%, 25-35 years:  24.0\%, 35-45 years:15.5\%, 45-55 years: 10.1\%, 55+ years:  28.7\%. Regarding gender, 75.2\% identify as female, 22.5\% as male, and 2.3\% as non-binary. In terms of education level, 51.9\% have a graduate degree, 38.8\% have a college degree, and 9.3\% have a high-school diploma. The racial composition is 81.4\% White, 0.8\% American Indian or Alaska Native, 11.6\% Asian, 2.3\% Black or African American, and 3.9\% prefer not to say. For first language, 93\% are English speakers and 7\% speak a non-English language. In terms of English proficiency, 77.5\% are native speakers, 21.7\% are fluent, and 0.8\% are not fluent. Regarding writing expertise, 5.4\% are professional writers, 62.8\% write regularly, and 31.8\% do not write regularly. For prior use of assistive technology in writing, 31\% have not used any, 58.9\% have used basic features like spell check, and 10.1\% have used advanced writing assistants. Lastly, attitudes toward working with AI tools are as follows: 17.8\% enjoy it, 51.9\% would like to collaborate, and 30.3\% find it fun.}
  \label{table:demographic_breakdown}
\end{table*}

\subsection{Data Collection and Measurements}

We collected several pieces of information from the participants in our experiment.

\begin{itemize}
    \item {\it Basic Demographic Information:} At the beginning of the study, we asked participants to self-report critical variables such as Age, Gender, Ethnicity, Race, and Level of Education. The purpose was to understand the demographic makeup of our participant pool, which is crucial for interpreting the generalizability of our findings.
    \item {\it English and Writing Proficiency:} Questions were asked to gauge participants' English and writing proficiency. Specifically, participants were asked about their First Language, Level of English Proficiency (Native Speaker, Fluent, Not Fluent), and Level of Writing Expertise (Professional Writer, Writes Regularly, Does Not Write Regularly). These measures are designed to understand the extent to which language and writing skills influence interaction with the AI tool.
    \item {\it Prior AI Experience and Attitude:} To explore participants' prior experience and attitudes toward AI, we inquired about their Prior Use of Assistive Technology in Writing (No, Basic Usage, Advanced Writing Assistants) and their Attitude Toward Working with AI Tools (Enjoying working with AI, Would Like to Collaborate with AI, Finds it Fun to Work with AI). This information helps to provide context for how predisposed participants are to accepting or rejecting AI suggestions.

\Y{However, it is important to note that the survey question on AI attitudes was not a criterion for participant selection, nor was it used in our analyses. This aspect of the survey served solely as additional context and did not influence the outcomes of the study. We acknowledge the inherent challenge in recruiting participants with neutral or negative views towards AI within our research context. Future studies could aim to capture a broader range of attitudes, including those less favorable towards AI, to garner more varied and possibly insightful perspectives.}

    \item {\it Writing Results from Participants:} We gathered complete sets of final text submissions from participants for each writing prompt which were later evaluated for text quality. This unprocessed information serves as a foundational element for evaluating the effectiveness of collaborations between humans and AI in writing tasks.
    \item {\it Writing Interaction Log Data:} We measured various metrics to capture the nature of participants' interactions with the AI tool. These included all the user actions (e.g. insert, [text1], delete, [text2]), the number of edits, the count of words with spelling errors, and the time used for each task, which are engineered to facilitate subsequent processing and evaluation.
    \item {\it Post-task Self-Evaluations:} Participants completed a survey to self-evaluate their perceived cognitive and emotional workload after completing each writing task, utilizing the NASA Task Load Index (NASA-TLX)~\cite{HART1988139} (described in Table~\ref{table:postsurvey-questions}). Measures included Mental Demand, Physical Demand, Temporal Demand, Effort, Performance, and Frustration. These scales ranged from low to high or from good to poor. 
    \item {\it Post-task User Experience of AI Collaboration:}
    We also explored participants' experience collaborating with AI in the post-task survey (described in Table~\ref{table:postsurvey-questions}). The questions included user satisfaction with the output, ownership feelings, AI influence on their writing, task difficulty, and future AI use likelihood. These aspects were also measured on scales from low to high.
\end{itemize}

\section{Empirical Analysis}
After collecting the data from N=131 participants, we proceeded to analyze the data. We performed a series of quantitative analyses using several statistical models as discussed below.

\subsection{Outcome variables:}
We considered several broad categories of outcomes variables.

{\noindent \bf Output Quality:}
    \begin{itemize}
        \item {\it Text Quality:} 
        Quantifying the quality of written text is challenging due to the lack of a universally accepted definition. To address this, we utilized three different metrics for assessing text quality. We adopted two automated text analysis tools, TAACO\footnote{\url{https://www.linguisticanalysistools.org/taaco.html}} and TAALES\footnote{\url{https://www.linguisticanalysistools.org/taales.html}}, which are widely recognized for analyzing the linguistic structure of texts, specifically focusing on text cohesion (TAACO) and lexical sophistication (TAALES)\textemdash key indicators of text quality. Additionally, we incorporated a third metric based on independent human evaluations by two independent raters, assessing the writing based on its response to the prompt, the conciseness of the answer, and the quality of argumentation. Given the strong correlation between our human evaluation metrics and the TAACO and TAALES scores, we chose to report our findings primarily using the human evaluation as the measure of text quality in this study. All three quality scores were on a 1-7 scale.
        \item {\it Number of Edits:} The total number of edits (additions, deletions, changes) made by the participant.
        \item {\it Number of Errors:} The total number of spelling errors in the text output.
    \end{itemize}
    {\noindent \bf Emotional \& Cognitive Investment:}
     \begin{itemize}
        \item {\it User Satisfaction:} How satisfied is the user with the output of collaboration (1-20 scale)?
        \item {\it Content Ownership:} How much ownership does the user feel over the output of collaboration (1-20 scale)?
        \item {\it NASA Cognitive Load Index:} The average of NASA's Task Load Indices~\cite{HART1988139} (1-20 scale): mental demand, physical demand, temporal demand, effort, performance, and frustration.
    \end{itemize}
    {\noindent \bf Task Efficiency \& Persuasion:}
     \begin{itemize}
        \item {\it Productivity (words/time):} Total number of words written per unit time spent writing the response.
          \item {\it Influence:} How much of the user response was influenced by the AI text suggestions (1-20 scale)?
    \end{itemize}

\subsection{Model Details:}
We performed several empirical analyses on the data that we collected.

\begin{table*}[ht]
\centering
\label{tab:glm_results_extended_aic}
\begin{tabular}{lccc}
\hline
Coefficient & Edits & Errors & Text Quality \\
\hline
Intercept & \textbf{2462.48*** (115.03)} & \textbf{30.36*** (1.30)} & \textbf{3.70*** (0.16)} \\
$condition_j$: paragraph & \textbf{-795.04*** (92.86)} & 0.35 (1.05) & \textbf{0.18* (0.06)} \\
$condition_j$: sentence & \textbf{-334.14*** (92.64)} & -1.08 (1.05) & \textbf{-0.29* (0.13)} \\
$position_{ij}$ & -53.25 (46.58) & 0.49 (0.53) & 0.09 (0.06) \\
\hline
\hline
\end{tabular}
\caption{Model 1 estimation results quantifying the impact of AI scaffolding on \underline{Output Quality.} The ``no AI support'' condition is the reference condition. We estimate a Generalized Linear Model (GLM) model with Gaussian family. We correct for multiple comparisons via False Discovery Rate (FDR)~\cite{benjamini1995controlling}. {Note:} N=393, Significance codes: 0 ‘***’ 0.001 ‘**’ 0.01 ‘*’ 0.05 ‘.’ 0.1 ‘ ’ 1.}
\end{table*}

\subsubsection{\bf Quantifying the effect of AI scaffolding on the writing outcomes}
\begin{itemize}
\item {\bf Model 1:} $Y_{ij} =\beta_0+ \beta_1*condition_j + \beta_2*position_{ij} + \epsilon_{ij}$
\item In this statistical model, $Y_{ij}$ represents various outcome measures described earlier. They included output quality, emotional and cognitive investment, task efficiency, and persuasion variables for the $i^{th}$ participant in the $j^{th}$ condition.\footnote{The `output quality' outcome variables are our primary outcomes of interest, so our results will focus more on them.} $\beta_0$ denotes the intercept term, $condition_j$ encodes the dummy variables for different conditions, and $position_{ij}$ represents the sequence position at which the $i^{th}$ participant encountered the $j^{th}$ condition.

\item We use a standard generalized linear model (GLM) to analyze our data, a typical approach for experimental designs like the one employed in this study. We estimated a separate regression model for each outcome variable. Since we have multiple observations per subject, we cluster the standard errors at the level of users, to control for serial correlation of errors. Furthermore, given that we test hypotheses across multiple outcome variables, we face the potential for false positives arising from multiple comparisons. To mitigate this risk, we adjusted our treatment p-value estimates by applying a False Discovery Rate (FDR) correction~\cite{benjamini1995controlling}.

\item Our primary focus is on evaluating the influence of the $condition_j$ dummy variable (baseline, sentence, paragraph) on the outcomes. We also control for $position_{ij}$, which indicates the sequence position at which a specific condition was encountered by a participant. For example, the baseline condition (no AI assistance) occurs at position 1 for users in sequence 1, position 2 for those in sequence 2, and position 3 for users in sequence 3.
\end{itemize}

\subsubsection{\bf Quantifying the differential impact of AI scaffolding on different users}
\begin{itemize}
\item {\bf Model 2:} $Y_{ij} = \beta_0 +\beta_1*expertise_i + \beta_{3}*condition_j\times expertise_i + \beta_{2}*position_{ij} + \epsilon_{ij}$
\item {\bf Model 3:} $Y_{ij} = \beta_0 +\beta_1*tech\_use_i + \beta_{3}*condition_j\times tech\_use_i + \beta_{2}*position_{ij} + \epsilon_{ij}$
\item In both these models, the variable definitions $Y_{ij}$, $condition_j$, $position_{ij}$ for the $i^{th}$ participant in the $j^{th}$ condition are the same as earlier, but we only focus on {\it Text Quality} and {\it Errors} as the key outcome variables to study here. $expertise_i$ codes the writing proficiency (not regular, regular, and proficient) of the user and $tech\_use_i$ represents their tech-savviness (no technology use, basic use, and advanced use). Both these are coded as dummy variables.
\item The interaction term of $expertise_i$ and $tech\_use_i$ with $condition_j$ helps us quantify the differential impact of these variables for the various scaffolding conditions (baseline, sentence, paragraph).
\item Once again, we estimate a GLM model and cluster the standard errors at the level of users.
\end{itemize}

\subsubsection{\bf Quantifying the positional impact of AI scaffolding on different users}
\begin{itemize}
\item {\bf Model 4:} $Y_{ij} = \beta_0 + \beta_{1}*previous\_condition_j + \beta_2*position_{ij} + \epsilon_{ij}$
\item $Y_{ij}$ is the {\it Text Quality}, $previous\_condition$ is the last condition in the sequence that the user encountered (none, baseline, sentence, paragraph). And, as earlier, $position_{ij}$ represents the sequence position at which the $i^{th}$ participant encountered the $j^{th}$ condition.

\item Consistent with our prior analyses, we employed a standard GLM model and cluster the standard errors at the level of users.
\end{itemize}

\section{Results}

\subsection{How did the ``Output Quality'' change with scaffolding?}
First, our results (Table 3) reveal a substantial reduction in the number of edits made by users in both low-scaffolding (sentence-level suggestions) and high-scaffolding (paragraph-level suggestions) conditions, with reductions of -334.1 (p<0.0001) and -795.04 (p<0.0001), respectively. This outcome aligns with the intuitive understanding that in the baseline condition, where no AI assistance is provided, users have the task of crafting the text from scratch. Conversely, in AI-scaffolded scenarios, users predominantly focus on refining the pre-generated text, thereby requiring fewer edits.

Second, an unexpected observation is the lack of a statistically significant difference in the number of spelling errors across all conditions. Despite AI-scaffolded conditions supplying users with an initial draft, the frequency of spelling errors remains consistent with that of the baseline condition. This suggests that the presence of AI assistance doesn't noticeably impact the types of errors made during the editing process.

Finally, a noteworthy finding is the differential impact of low and high scaffolding on writing quality. The quality of writing in the sentence-level suggestion mode deteriorates compared to the baseline, indicated by a decline of -0.29 quality points (p=0.02). In contrast, the paragraph-level suggestion mode led to a significant enhancement in writing quality by 0.18 quality points (p=0.02).

This U-shaped pattern in writing quality can be interpreted as follows: Sentence-level suggestions may inadvertently hamper users' ability to think holistically about their responses. Instead of aiding in the thought process, these isolated suggestions might contribute to a fragmented narrative, leading to a drop in the overall quality of the output text. On the other hand, paragraph-level suggestions offer a more comprehensive framework, which appears to help users in producing higher-quality text. 

\begin{table*}[ht]
\centering
\begin{tabular}{lccccc}
\hline
Coefficient & Satisfaction & Ownership & NASA Index & Influence & Productivity (words/time) \\
\hline
Intercept & \textbf{13.73*** (0.84)} & \textbf{19.39*** (0.64)} & \textbf{8.53*** (0.58)} & -0.40 (0.61) & \textbf{0.34*** (0.03)} \\
$condition_j$: Paragraph & \textbf{-1.89** (0.68)} & \textbf{-5.65*** (0.54)} & -0.66 (0.47) & \textbf{5.91*** (0.49)} & \textbf{0.07** (0.02)} \\
$condition_j$: Sentence & \textbf{-2.14** (0.66)} & \textbf{-3.66*** (0.51)} & -0.22 (0.43) & \textbf{5.07*** (0.46)} & -0.02 (0.03) \\
$position_{ij}$ & -0.15 (0.34) & -0.02 (0.26) & -0.09 (0.23) & 0.20 (0.25) & \textbf{0.03* (0.01)} \\
\hline
\hline
\end{tabular}
\caption{Model 1 estimation results quantifying the impact of AI scaffolding on \underline{Emotional \& Cognitive Investment} and \underline{Task Efficiency \& Persuasion} variables. The ``no AI support'' condition is the reference condition. We estimate a Generalized Linear Model (GLM) model with Gaussian family. We correct for multiple comparisons via False Discovery Rate (FDR)~\cite{benjamini1995controlling}. {Note:} N=393, Significance codes: 0 ‘***’ 0.001 ‘**’ 0.01 ‘*’ 0.05 ‘.’ 0.1 ‘ ’ 1.}
\label{tab:glm_results_combined_metrics}
\end{table*}

\subsection{How did the ``Emotional \& Cognitive Investment'' change with scaffolding?}
As can be seen from the results in Table~\ref{tab:glm_results_combined_metrics}, satisfaction was lower in both the sentence and paragraph scaffolding conditions compared to the control, with sentence scaffolding (-2.14, p=0.001) resulting in greater dissatisfaction than paragraph scaffolding (-1.89, p=0.001), even though paragraph scaffolding led to higher text quality. This suggests a paradox where higher quality (in the case of paragraph mode) does not necessarily increase satisfaction. One explanation could be ``effort justification,'' where satisfaction is tied to the perceived effort. In paragraph scaffolding, the provided drafts might make users feel they've exerted less effort, lowering satisfaction despite better quality~\cite{norton2012ikea, de2016almost}. This indicates users may value the writing process as much as the outcome~\cite{card2020referees}.

Similarly, the sense of ownership over the generated text was found to be significantly diminished in both AI-assisted conditions. The loss of ownership was more pronounced in the paragraph condition, at -5.65 (p<0.0001), compared to -3.65 (p<0.0001) in the sentence condition. This may indicate that while the paragraph-level suggestions improve the final text's quality, they also potentially make the user feel less involved or invested in the writing process.

Remarkably, no significant changes were observed in the cognitive loads among the three conditions, as measured by the NASA load index. This could imply that while the level of scaffolding impacts user satisfaction and sense of ownership, it does not substantially affect the cognitive effort required to complete the writing task.

\subsection{How did the ``Task Efficiency \& Persuasion'' change with scaffolding?}
The results (Table~\ref{tab:glm_results_combined_metrics})  indicate that the impact of AI suggestions on the final written product is significantly higher for users in the paragraph condition, registering a score of 5.91 (p<0.0001). This is somewhat higher than the influence observed in the sentence condition, which had a score of 5.07 (p<0.0001). This observation aligns well with a recent study by Jakesch et al.~\cite{Jakesch2023}, which also highlighted the role of AI in shaping users' decisions and opinions.

Furthermore, the paragraph condition had a notably positive effect on productivity, increasing the rate of writing by an additional 0.07 words per unit time (p=0.01). Interestingly, the sentence condition did not demonstrate a significant productivity advantage over the baseline. This might suggest that while paragraph-level suggestions provide enough content to boost writing speed, sentence-level suggestions may not offer sufficient guidance to make a noticeable difference in productivity.

\subsection{Did the impact of AI scaffolding vary with Users' Writing Expertise and Prior Technology Usage?}
The participants in our study had three different levels of writing expertise\textemdash not regular (i.e., the user does not write regularly), regular (i.e., user writes regularly), \& professional writers (i.e., user is a professional writer). The AI conditions slightly increased the number of errors made by both regular and professional writers (around 3 more errors) compared to the baseline no AI assistance condition (Table~\ref{tab:glm_results_expertise_actual}). This trend may imply that when using AI-generated suggestions, users may focus more on shaping the overall flow of the text and neglect to catch small mistakes, effectively going into an ``auto-pilot'' mode~\cite{Yang2022,schuster2021get}.

\begin{table*}[ht]
\centering
\begin{tabular}{lcc}
\hline
Coefficient & Errors & Text Quality \\
\hline
Intercept & \textbf{29.86*** (1.72)} & \textbf{3.45*** (0.20)} \\
$position_{ij}$ & 0.47 (0.53) & 0.11 (0.06) \\
$expertise_i$: Professional & \textbf{3.25*** (0.45)} & 1.25 (1.41) \\
$expertise_i$: Regular & \textbf{2.56* (1.16)} & \textbf{0.26* (0.02)} \\
$expertise_{i}$: Not Regular $\times$ $condition_j$: Paragraph & -2.15 (1.83) & \textbf{0.53* (0.22)} \\
$expertise_{i}$: Professional $\times$ $condition_j$: Paragraph & -2.48 (4.52) & -0.65 (0.54) \\
$expertise_{i}$: Regular $\times$ $condition_j$: Paragraph & 1.92 (1.32) & \textbf{0.07* (0.02)} \\
$expertise_{i}$: Not Regular $\times$ $condition_j$: Sentence & -1.32 (1.82) & -0.13 (0.22) \\
$expertise_{i}$: Professional $\times$ $condition_j$: Sentence & -1.41 (4.51) & -0.78 (0.54) \\
$expertise_{i}$: Regular $\times$ $condition_j$: Sentence & -0.93 (1.32) & \textbf{-0.33* (0.16)} \\
\hline
\hline
\end{tabular}
\caption{Model 2 estimation results quantifying the differential impact of AI scaffolding on \underline{Errors} and  \underline{Text Quality} variables stratified according to user writing expertise. The ``no AI support'' condition is the reference condition. We estimate a Generalized Linear Model (GLM) model with Gaussian family. {Note:} N=393, Significance codes: 0 ‘***’ 0.001 ‘**’ 0.01 ‘*’ 0.05 ‘.’ 0.1 ‘ ’ 1.}
\label{tab:glm_results_expertise_actual}
\end{table*}

On a more optimistic note, AI-driven paragraph-level suggestions significantly improved text quality for non-regular and regular writers. For non-regular writers, the quality score saw an average boost of 0.53 (p-value=0.03). For regular writers, there was a slight but statistically significant quality improvement of 0.07 (p-value=0.02). In contrast, AI interventions had no statistically significant impact on the writing quality of professional writers~\cite{buschek2021impact}. This could suggest that professional writers already possess well-honed skills that AI cannot readily augment~\cite{gero2019stylistic}.

Participants also varied in technology usage: none, basic, advanced. No significant error variation was found across these groups (see Table 6). However, paragraph-level AI positively affected users with no technology expertise (increase of 0.55, p=0.02) and sentence-level AI decreased text quality for those with basic technology expertise (decrease of -0.38, p=0.03), highlighting the nuanced impacts of AI assistance based on technology proficiency.

\begin{table*}[ht]
\centering
\label{tab:glm_results_technology_final}
\begin{tabular}{lcc}
\hline
Coefficient & Errors & Text Quality \\
\hline
Intercept & \textbf{31.44*** (2.54)} & \textbf{3.68*** (0.30)} \\
$position_{ij}$ & 0.43 (0.53) & 0.09 (0.06) \\
$tech\_use_i$: Basic & -1.53 (2.54) & 0.23 (0.30) \\
$tech\_use_i$: No & -0.19 (2.71) & -0.37 (0.32) \\
$tech\_use_i$: Advanced $\times$ $condition_j$: Paragraph & 2.76 (3.33) & 0.20 (0.39) \\
$tech\_use_i$: Basic $\times$ $condition_j$: Paragraph & -0.25 (1.36) & -0.01 (0.16) \\
$tech\_use_i$: No $\times$ $condition_j$: Paragraph & 0.71 (1.90) & \textbf{0.55* (0.22)} \\
$tech\_use_i$: Advanced $\times$ $condition_j$: Sentence & -2.50 (3.32) & -0.55 (0.39) \\
$tech\_use_i$: Basic $\times$ $condition_j$: Sentence & -1.06 (1.36) & \textbf{-0.38* (0.16)} \\
$tech\_use_i$: No $\times$ $condition_j$: Sentence & -0.65 (1.89) & -0.02 (0.22) \\
\hline
\hline
\end{tabular}
\caption{Model 3 estimation results quantifying the differential impact of AI scaffolding on \underline{Errors} and  \underline{Text Quality} variables stratified according to user technology savviness. The ``no AI support'' condition is the reference condition. We estimate a Generalized Linear Model (GLM) model with Gaussian family. {Note:} N=393, Significance codes: 0 ‘***’ 0.001 ‘**’ 0.01 ‘*’ 0.05 ‘.’ 0.1 ‘ ’ 1.}
\end{table*}

\subsection{Did the Position in the Sequence or the Previous Condition affect AI Co-writing Quality?}
Our results (Table~\ref{tab:glm_results_quality}) also show a subtle link between the position of a {\it condition} in the sequence and text quality. Text quality slightly increased by 0.37 as participants progressed through conditions (p=0.01), suggesting familiarity with the interface improves performance~\cite{luger2016like}. Initial exposure to conditions seems crucial, with the highest text quality observed in the `first condition,' indicating participants' attentiveness impacts their output. Notably, transitioning from sentence to paragraph-level suggestions led to a significant quality boost of 0.52 (p<0.0001), hinting at a beneficial learning effect from prior sentence-level experience.

However, text quality dropped notably when transitioning from paragraph-level assistance back to the no AI baseline, implying a possible over-reliance on AI support~\cite{yin2019understanding}. This suggests that while AI can enhance writing initially, dependence on it might reduce effort and quality when AI assistance is withdrawn.

\begin{table}[ht]
\centering
\begin{small}
\begin{tabular}{lc}
\hline
Coefficient & Text Quality \\
\hline
Intercept & \textbf{2.73*** (0.34)} \\
$position_{ij}$ & \textbf{0.37** (0.13)} \\
$previous\_condition_j$: First Condition & \textbf{0.78** (0.24)} \\
$previous\_condition_j$: Paragraph & 0.06 (0.15) \\
$previous\_condition_j$: Sentence & \textbf{0.52*** (0.16)} \\
\hline
\hline
\end{tabular}
\end{small}
\caption{Model 4 estimation results quantifying the impact of previous AI support condition and the sequence position on \underline{Text Quality}. The ``no AI support'' condition is the reference condition. We estimate a Generalized Linear Model (GLM) model with Gaussian family. {Note:} N=393, Significance codes: 0 ‘***’ 0.001 ‘**’ 0.01 ‘*’ 0.05 ‘.’ 0.1 ‘ ’ 1.}
\label{tab:glm_results_quality}
\end{table}

\section{Discussion}
The collaborative writing process, often considered a quintessentially human endeavor, is undergoing a transformative shift with the advent of AI writing assistants. As this technological intervention gains traction, it becomes imperative to understand the nuanced interactions between human creativity and machine-driven guidance. Our research dives into this intriguing intersection, interrogating how different levels of AI scaffolding influence the quality of the written output and the user experience in the collaborative writing environment. Our results not only offer actionable insights for optimizing such hybrid writing systems but also touch on the psychological and pedagogical facets of human-AI interaction in a writing context. Below, we summarize the key findings, discuss their underlying mechanisms, and finally discuss our research's practical and ethical implications.

\subsection{Summary of Key Results and Potential Mechanisms}
Our work provides several key insights into the impact of varied levels of AI scaffolding on collaborative writing processes and outcomes. A prominent finding is the U-shaped relationship between the level of scaffolding and writing quality/productivity\textemdash while low scaffolding adversely affected these metrics, high scaffolding conferred significant benefits. {\it What mechanisms might underpin such trends?} 

A conjecture is that sentence-level AI suggestions, or low-level scaffolding, might disrupt writing by forcing writers to alternate between creating text and assessing disjointed AI inputs, lacking broader context. This could deteriorate the writing's quality and coherence. This issue aligns with the ideas presented by Pea~\cite{pea2018social}, who emphasized in his work on distributed intelligence and scaffolding that effective support should enhance and broaden a writer's cognitive abilities, not break them into smaller, disjointed parts.

On the other hand, paragraph-level suggestions provide broader, narrative-structuring guidance, helping writers integrate, adjust, or enhance the AI-generated content. This method reduces cognitive strain and aids in maintaining narrative coherence, fitting Vygotsky's Zone of Proximal Development (ZPD) concept~\cite{vygotsky1978development}, which suggests optimal learning occurs with challenges that are within the learner's (or writers, in this context) reach. By offering a structured yet adaptable framework, paragraph-level AI suggestions sit within this zone, enhancing the writer's capability to organize and develop ideas effectively.

However, the augmentation in writing experience with paragraph-level scaffolding is juxtaposed with a decline in user satisfaction and sense of ownership, reflecting Pea's~\cite{pea2018social} views on augmenting versus maintaining a user's agency. Extensive AI content could diminish the user's sense of effort and creativity, affecting their engagement and satisfaction. This highlights the importance of balancing efficiency and user experience in AI collaboration, in line with Vygotsky's Zone of Proximal Development (ZPD)~\cite{vygotsky1978development}, emphasizing the need for active user involvement in developing AI-based co-writing environments.

The sequence effects that we observe in this study reveal that moving from sentence-level to paragraph-level AI suggestions improves writing quality, suggesting a learning effect. This aligns with cognitive learning theories~\cite{piaget1952origins} that posit initial, simple interactions with a system can enhance later use of more sophisticated functions. Starting with sentence-level suggestions helps users get accustomed to the AI, aiding in better performance with paragraph-level suggestions later. Such a progression also resonates with the findings of the scaffolding concepts advanced by Wood et al.~\cite{wood1976role} and Vygotsky~\cite{vygotsky1978development}, where support is initially provided at a basic level and then progressively tailored to more complex tasks.

The decline in writing quality after removing AI scaffolding suggests an over-reliance on AI support, reflecting the `learned helplessness' concept in psychology~\cite{maier2016learned}, where individuals become accustomed to external assistance to the point that they neglect their own capabilities in its absence. As a result, users might depend too much on paragraph-level AI assistance, diminishing their use of creativity and critical thinking. This highlights the need for balance in AI assistance, as excessive reliance can undermine the development of independent writing skills.

In summary, the study's findings reveal promising opportunities for AI writing assistants to augment writing outcomes but also accentuate the importance of adaptive, personalized scaffolding approaches. The results advocate for the development of human-centered AI writing tools that not only enhance writing quality but also foster the growth of users' writing skills. Such tools should strategically balance the level of assistance they provide, ensuring that they augment rather than replace the human cognitive processes involved in writing.

\subsection{Implications for Human-AI Co-writing}
The insights from our study offer a nuanced perspective on designing future human-AI co-writing systems and mixed-initiative systems, underscoring the potential and challenges of integrating AI into the writing process.

One key aspect highlighted by our findings is the value of AI in providing structured frameworks for writing. High-level paragraph suggestions from AI can significantly aid users in organizing their thoughts and crafting comprehensive content. However, it's crucial to manage the extent of AI-generated content carefully. An overabundance of AI input can lead to a reduction in the writer's sense of involvement and creative expression. Future systems should, therefore, focus on a balanced approach, where AI contributes to the creative process without dominating it. This could involve AI tools that offer guidance or ideas while leaving ample space for the user’s creative input.

Our research shows the importance of tailoring AI assistance to user skill levels. Beginners may need more AI help, while experienced writers require less to maintain independence.  As users become more adept, the level of AI assistance should be dialed back, allowing for greater independence in writing. Future studies could investigate systems that adjust AI support according to a user’s evolving proficiency.

Personalization emerges as another crucial factor in the design of future co-writing systems. Our study shows that the impact of AI assistance differs across users with varying writing experiences. Less frequent writers benefit from more AI help, whereas experienced writers need less. Future systems should adapt AI support to individual preferences, using techniques to understand users’ backgrounds and needs.

In conclusion, our study advocates for a flexible, user-centered approach in the development of human-AI co-writing tools. AI systems should be designed to adapt to the evolving skills and preferences of users, enhancing the writing process without undermining the human element. By ensuring that AI scaffolding is responsive and sensitive to individual user needs, future co-writing systems can support and enhance human creativity and productivity in writing tasks. This approach aligns with the broader objectives in Human-Computer Interaction (HCI), where the goal is not just to create advanced technologies, but to develop tools that enrich human experiences and capabilities.

\subsection{Limitations and Generalizability}
While our study provided substantial insights into designing human-AI co-writing interfaces, our setup is not without limitations.

First, the experiment examined only two discrete levels of scaffolding—sentence and paragraph suggestions. Testing a wider range of scaffolding types including word-level, multi-paragraph, section-level, and dynamically adaptive systems would be an excellent direction for future research to reveal additional nuances in optimal scaffolding design.

Second, the prompts covered only argumentative writing in a standardized test format. Examining other genres such as narrative, descriptive, expository, persuasive, and technical writing through similar controlled experiments would be an excellent future research avenue to thoroughly investigate different user needs and AI collaboration dynamics across diverse writing contexts.

Third, the sample primarily comprised English speakers proficient in writing. Testing effects specifically for non-native English speakers and those still developing writing skills presents an excellent opportunity for impactful future work to offer enhanced insights into optimal, inclusive scaffolding for broader populations.

Fourth, only short-term effects were studied in brief, one-time sessions. Longitudinal studies tracking changes in writing processes, quality, and attitudes with extended AI tool usage over weeks or months would be an excellent complement to the current results and reveal valuable patterns about long-term implications of writing scaffolds.

Finally, our experimental setting differs substantially from real-world writing contexts like academia, journalism, fiction, and technical writing. In-depth ethnographic observations or large-scale field studies across diverse professional and educational domains present an excellent future direction to highlight highly complementary insights regarding real-world integration, adoption, and impact of AI writing assistants.

In summary, expanding this research through studies involving more diverse conditions, genres, populations, timeframes, and naturalistic settings would excellently build generalizability and provide a comprehensive understanding of human-AI co-writing. Nonetheless, the current results constitute an important first step toward informing effective writing scaffold design.

\subsection{Ethical Considerations}
As with any study involving human participants, ethical considerations must be carefully weighed. All participants provided informed consent, and identifiers were anonymized to protect privacy. Compensation was appropriately calibrated to acknowledge participants' time without exerting undue influence. The study underwent review by an institutional ethics board to ensure protections were in place.

Nonetheless, responsible deployment of AI writing assistants necessitates ongoing ethical vigilance. Risks like plagiarism must be mitigated by design. The balance between augmenting and replacing human effort must be monitored, avoiding harmful over-reliance on automation. Biases in training data could propagate unfair impacts, requiring proactive evaluation. Transparency around AI limitations is imperative. Ultimately, human well-being and dignity should remain the highest priority as intelligent writing tools proliferate. Further interdisciplinary deliberation on ethics would guide the conscientious advancement of human-AI writing collaboration.

\begin{acks}
We would like to thank Alain Cohn, Tanya Rosenblat, Laura Aull, and four anonymous reviewers for constructive feedback that helped improve the paper.
\end{acks}
% Bib
\bibliographystyle{ACM-Reference-Format}
\bibliography{bib_file}

\appendix

\section{Pre-Screening Task Details}
\label{appendix:Pre_Screening_Task_Details}
\subsection{Pre-screening HAI Collaborative Writing Task}
\label{appendix:Pre_screening_HAI_Collaborative_Writing_Task}

{\bf Directions:} Give yourself 3 minutes to read expert 1’s opinion. 
{\bf Reading Time:} 3 minutes 

In an effort to encourage ecologically sustainable forestry practices, an international organization started issuing certifications to wood companies that meet high ecological standards by conserving resources and recycling materials. Companies that receive this certification can attract customers by advertising their products as “eco-certified.” Around the world, many wood companies have adopted new, ecologically friendly practices in order to receive eco-certification. However, it is unlikely that wood companies in the United States will do the same, for several reasons. 
First, American consumers are exposed to so much advertising that they would not value or even pay attention to the eco-certification label. Because so many mediocre products are labeled “new” or “improved,” American consumers do not place much trust in advertising claims in general. 
Second, eco-certified wood will be more expensive than uncertified wood because in order to earn eco-certification, a wood company must pay to have its business examined by a certification agency. This additional cost gets passed on to consumers. American consumers tend to be strongly motivated by price, and therefore they are likely to choose cheaper uncertified wood products. Accordingly, American wood companies will prefer to keep their prices low rather than obtain eco-certification. 
Third, although some people claim that it always makes good business sense for American companies to keep up with the developments in the rest of the world, this argument is not convincing. Pursuing certification would make sense for American wood companies only if they marketed most of their products abroad. But that is not the case—American wood businesses sell most of their products in the United States, catering to a very large customer base that is satisfied with the merchandise.

{\bf Directions:} Now, give yourself 3 minutes to read the expert 2’s opinion:

Well, despite what many people say, there’s good reason to think that many American wood companies will eventually seek eco-certification for their wood products. First off, consumers in the United States don’t treat all advertising the same. They distinguish between advertising claims that companies make about their own products and claims made by independent certification agencies. Americans have a lot of confidence in independent consumer agencies. Thus, ecologically minded Americans are likely to react very favorably to wood products ecologically certified by an independent organization with an international reputation for trustworthiness. 
Second point—of course it’s true that American consumers care a lot about price—who doesn’t? But studies of how consumers make decisions show that price alone determines consumers’ decisions only when the price of one competing product is much higher or lower than another. When the price difference between two products is small—say, less than five percent, as is the case with certified wood— Americans often do choose on factors other than price. And Americans are becoming increasingly convinced of the value of preserving and protecting the environment. 
And third, U.S. wood companies should definitely pay attention to what’s going on in the wood business internationally, not because of foreign consumers, but because of foreign competition. As I just told you, there’s a good chance that many American consumers will be interested in eco-certified products. And guess what, if American companies are slow capturing those customers, you can be sure that foreign companies will soon start crowding into the American market, offering eco-certified wood that domestic companies don’t.

{\bf Task:} Summarize the points made in the text expressing the second expert’s opinion, being sure to explain how they cast doubt on specific points made in the text expressing the first expert’s opinion.
Your response is judged on the quality of the writing and on how well it presents the points in the text for the second expert's opinion and their relationship to the first expert's opinion. Typically, an effective response will be 150 to 225 words. 

{\bf Response time:} 20 minutes.

\subsection{Pre-screening Task Rubric}
\label{appendix:Pre_screening_Task_Rubric}

In the following rubric anywhere we say “lecture” we are referring to “expert2’s opinion” in the pre-screening task
Anywhere we say “reading”, we are referring to “expert 1’s opinion” in the pre-screening task.

\begin{table*}[H]
  \caption{Pre-screening Task Rubric}
  \label{table:Pre-screening_Rubric}
  \begin{tabular}{lp{6cm}p{6cm}}
    \toprule
    & Point made in the reading & Counterpoint made in the lecture \\
    \midrule
    1 &Because American consumers have come to distrust frequently used advertising claims such as ‘new’ or ‘improved,’ they won’t pay attention to or trust the eco certified label. & American consumers do pay attention to claims about products when those claims are made by independent consumer agencies. \\
    
    2 & Since eco certification adds to the cost of a product, Americans would be unlikely to buy eco certified products and would choose cheaper, uncertified products. & This is true only if there is a big price difference between two similar products; if an eco certified product costs only about five percent more, American consumers would accept this in order to buy the product that is better for the environment. \\
    
    3 & Because American companies sell their products mainly in the U.S., they do not need to compete in the rest of the world where eco certification is desired by consumers. & American companies must be ready to compete with foreign companies that will soon be selling eco certified products in the U.S. market. \\
    
    \bottomrule
  \end{tabular}
\end{table*}

\begin{enumerate}
 
   \item  {\bf Score 5:} A response at this level successfully selects the important information from the lecture and coherently and accurately presents this information in relation to the relevant information presented in the reading. The response is well organized, and occasional language errors that are present do not result in inaccurate or imprecise presentation of content or connections.

    \item  {\bf Score 4:} A response at this level is generally good in selecting the important information from the lecture and in coherently and accurately presenting this information in relation to the relevant information in the reading, but it may have minor omissions, inaccuracy, vagueness, or imprecision of some content from the lecture or in connection to points made in the reading. A response is also scored at this level if it has more frequent or noticeable minor language errors, as long as such usage and grammatical structures do not result in anything more than an occasional lapse of clarity or in the connection of ideas. 

   \item  {\bf Score 3:} A response at this level contains some important information from the lecture and conveys some relevant connection to the reading, but it is marked by one or more of the following: 
   \begin{itemize}
\item Although the overall response is definitely oriented to the task, it conveys only vague, global, unclear, or somewhat imprecise connection of the points made in the lecture to points made in the reading. 
\item The response may omit one major key point made in the lecture. 
\item Some key points made in the lecture or the reading, or connections between the two, may be incomplete, inaccurate, or imprecise. 
\item Errors of usage and/or grammar may be more frequent or may result in noticeably vague expressions or obscured meanings in conveying ideas and connections. 
\end{itemize}

   \item  {\bf Score 2:} A response at this level contains some relevant information from the lecture, but is marked by significant language difficulties or by significant omission or inaccuracy of important ideas from the lecture or in the connections between the lecture and the reading; a response at this level is marked by one or more of the following:
\begin{itemize}
\item The response significantly misrepresents or completely omits the overall connection between the lecture and the reading. 
\item The response significantly omits or significantly misrepresents important points made in the lecture. 
\item The response contains language errors or expressions that largely obscure connections or meaning at key junctures or that would likely obscure understanding of key ideas for a reader not already familiar with the reading and the lecture. 
\end{itemize}

   \item  {\bf Score 1:}  A response at this level is marked by one or more of the following: 
\begin{itemize}
 \item  The response provides little or no meaningful or relevant coherent content from the lecture. 
 \item  The language level of the response is so low that it is difficult to derive meaning. 
   
\end{itemize}

Score 0: A response at this level merely copies sentences from the reading, rejects the topic or is otherwise not connected to the topic, is written in a foreign language, consists of keystroke characters, or is blank.

\end{enumerate}

For the specific task we used in our pre-screening task, here is how to grade the writing:

What is important to understand from the lecture is that the professor disagrees with the points made in the reading, namely that American consumers mistrust advertising, that they are unwilling to pay extra for eco certified products, and that American companies do not need to compete in parts of the world where eco certification is valued. In your response, you should convey the reasons presented by the professor for why eco certification of wood should be adopted by U.S. companies. A high-scoring response will include the following points made by the professor that cast doubt on the points made in the reading.

\section{The custom-build interface: Surveys and Interaction Patterns}
\label{appendix:survey_interface}

\begin{figure*}[htbp]
  \centering
  \includegraphics[width=0.8\textwidth]{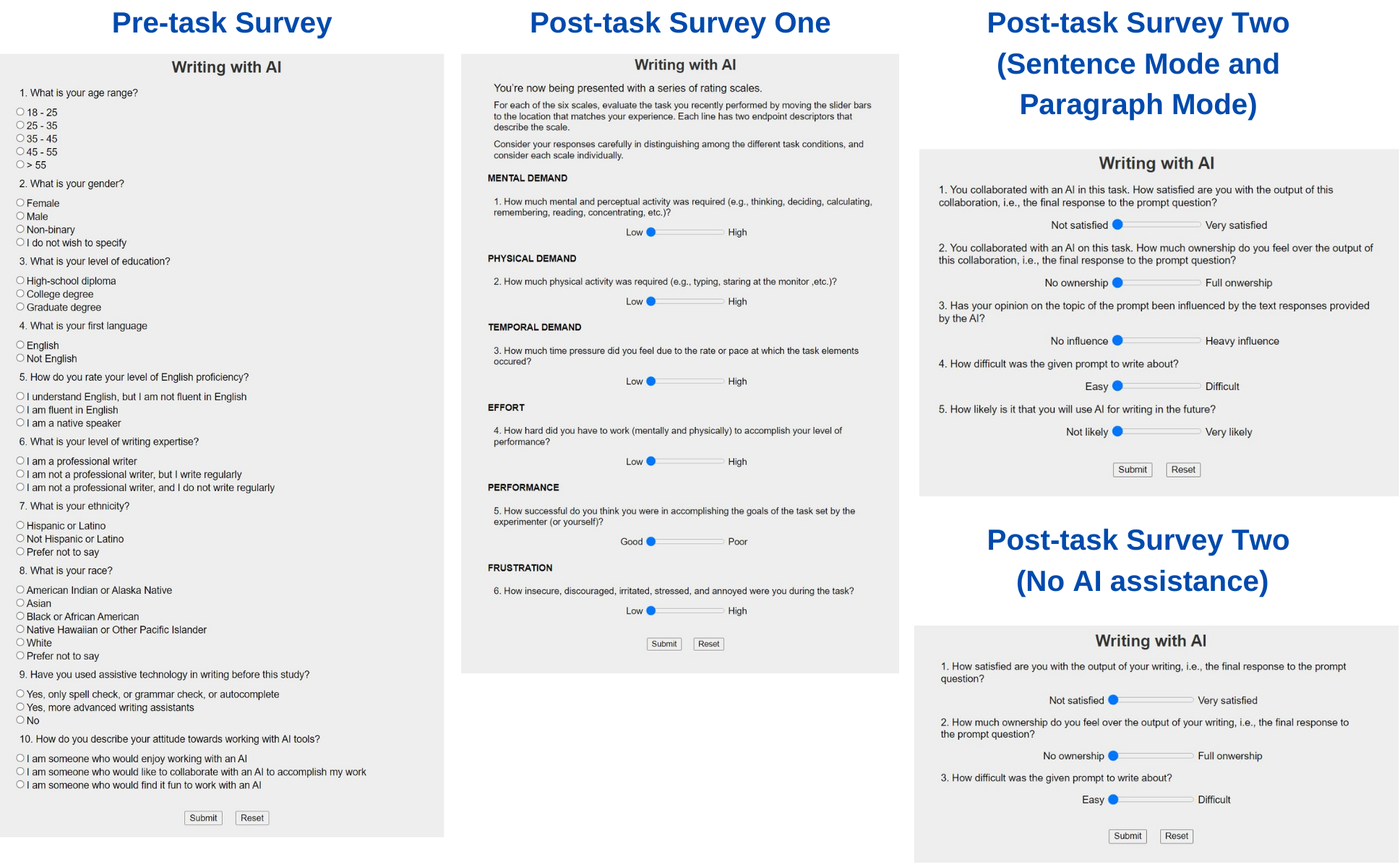} 
  \caption{The Custom-build Interface: Pre-task and Post-task Surveys}
  \Description{Figure 3 showcases four screenshots of our custom-built interface featuring Pre-task and Post-task Surveys. The far-left screenshot displays the Pre-task Survey, designed to collect demographic information and various characteristics such as age, gender, education level, English proficiency, writing expertise, and attitudes toward AI tools. The middle screenshot shows the first part of the Post-task Survey, which utilizes the NASA Task Load Index (NASA-TLX) to measure participants' perceived cognitive and emotional workload across six dimensions: Mental Demand, Physical Demand, Temporal Demand, Effort, Performance, and Frustration. Each dimension has a sliding scale for selection. The rightmost screenshots display the second part of the Post-task Survey, divided into two scenarios (Sentence and Paragraph Modes at the top, and No AI Assistance at the bottom). This part includes questions on Satisfaction, Ownership, Influence, Difficulty, and Likelihood of Future AI Use. These questions also feature sliding scales for selection. Submit and Reset buttons are present at the bottom of each survey interface.}
  \label{fig:surveys_interface}
\end{figure*}

Figure~\ref{fig:surveys_interface} shows the pre-task and post-task surveys administered to the participants. Table~\ref{table:action_description} shows the UI interaction details and controls for the AI scaffolded sentence and paragraph mode conditions.

\begin{table*}[htbp]
  \begin{small}
  \begin{tabular}{lllp{5cm}}
    \toprule
    Action & Triggered By & Control Keys & Description \\
    \midrule
    Request for Suggestions & User & Tab key & Initiate a query to the AI for a list of five suggestions, including the prompt and existing text. \\
    Show Suggestions & API & - & Render the list of five AI-generated suggestions on the user's screen. \\
    Navigate Suggestions & User & Arrow Keys \( \uparrow \downarrow \), Mouse & Allow users to scroll through the available suggestions. \\
    Choose Suggestion & User & Enter key; Mouse click & Select the suggestion. \\
    Dismiss Suggestions & User & Escape key & Remove the displayed suggestions and optionally request new ones. \\
    Modify Chosen Suggestion & User & Any key for editing, Mouse & Edit, or partially adopt the selected suggestion. \\
    \bottomrule
  \end{tabular}
  \end{small}
   \caption{User-AI Interaction Actions in Sentence and Paragraph Modes}
  \label{table:action_description}
\end{table*}

\section{Main writing study argumentative prompts}
\label{appendix:main_study_argumentative_prompts}
In our study, we employed a set of 10 argumentative writing prompts, carefully chosen to cover a wide range of subjects. These prompts were derived from an extensive collection utilized in the Co-Author paper~\cite{Lee_2022}.

\begin{itemize}
    \item \textbf{Screen:} How Worried Should We Be About Screen Time During the Pandemic? The coronavirus pandemic ended the screen time debate: Screens won. We all now find ourselves on our screens for school, for work and for connecting with family and friends during this time of social distancing and increased isolation. But should we be worried about this excessive screen use right now? Or should we finally get over it and embrace the benefits of our digital devices?
    \item \textbf{Dating:} How Do You Think Technology Affects Dating? Have you had any experience with dating? Have you ever used dating apps? If so, what has it been like for you? If not, why not? How do you think technology — like apps, Netflix, social media, and texting — affects dating and relationships? In your opinion, does it improve or worsen romantic interactions? How so?
    \item \textbf{Pads:} Should Schools Provide Free Pads and Tampons? Have you ever experienced period shaming, or ``period poverty?'' Should schools step in to help? Should schools be required to provide free pads and tampons to students? How are pads and tampons similar to toilet paper, soap, Band-Aids, and other products that are already provided in schools? How are they different?
    \item \textbf{School:} What Are the Most Important Things Students Should Learn in School? In your opinion, what are the most important things students should learn in school? What is the most important thing you have learned in school? How has this knowledge affected your life? How do you think it will help your success in the future?
    \item \textbf{Stereotype:} What Stereotypical Characters Make You Cringe? What stereotypical characters in books, movies, or television shows make you cringe and why? Would you ever not watch or read something because of its offensive portrayal of someone?
    \item \textbf{Audiobook:} Is Listening to a Book Just as Good as Reading It? Do you listen to audiobooks? What are the benefits, in your opinion, of listening instead of reading? Are there advantages to reading that cannot be gained by listening? Which method do you prefer? Why?
    \item \textbf{Athletes:} Should College Athletes Be Paid? Do you think college athletes should be paid? Or is a college scholarship and other non-monetary perks like the opportunity to play in front of cheering fans enough? What possible difficulties or downsides might there be in providing monetary compensation to players?
    \item \textbf{Extreme Sports:} Is It Selfish to Pursue Risky Sports Like Extreme Mountain Climbing? Some sports, like extreme mountain climbing, are dangerous. Since there are varying degrees of risk in most, if not all, sports (such as the possibility of concussions, broken bones, and even death), how does one decide where the line might be drawn between what is reasonable and what is not? Are some sports simply too dangerous to be called a sport?
    \item \textbf{Animal:} Is It Wrong to Focus on Animal Welfare When Humans Are Suffering? Would you be surprised to hear that a study found that research subjects were more upset by stories of a dog beaten by a baseball bat than of an adult similarly beaten? Or that other researchers found that if forced to choose, 40 percent of people would save their pet dog over a foreign tourist. Why do you think many people are more empathetic toward the suffering of animals than that of people? In your opinion, is it wrong to focus on animal welfare when humans are suffering? Why do you think so?
    \item \textbf{News:} Are We Being Bad Citizens If We Don’t Keep Up With the News? In your opinion, are we being bad citizens if we don’t keep up with the news? Do you think all people have some responsibility to know what is going on in the world? Does engaging with current events actually do anything at all? Why do you think the way you do?
\end{itemize}

\section{GPT Parameters}
\begin{table}[htbp]
  \begin{small}
  \begin{tabular}{l|c} 
    \toprule
    Parameter & Value \\
    \midrule
    model & text-davinci-002 \\
    max tokens & 30 (in sentence mode) or 150 (in paragraph mode) \\
    n & 5 \\
    temperature & 0.75 \\
    top p & 1 \\
    frequency penalty & 0.5 \\
    presence penalty& 0 \\
    \bottomrule
  \end{tabular}
  \end{small}
   \caption{Parameters used in API call. {\it Note:} n=Number of responses/completions requested, temperature=The temperature parameter is set between 0 and 1, with 0 being the most predictable and 1 being the most random response. We chose a value of 0.75 which is typically the recommended value for writing tasks where we want to see interesting variations in the generated text, frequency penalty=This parameter is used to discourage the model from repeating the same words or phrases too frequently within the generated text, presence penalty=This parameter is used to encourage the model to include a diverse range of tokens in the generated text.}
  \label{table:model_parameters}
\end{table}

We used GPT-3 davinci as the backend Large Language Model (LLM) for our experiments since it was the state-of-the-art when we performed this study. The parameters settings of GPT-3 that we used are shown in Table~\ref{table:model_parameters}.

\end{document}